# Spatial Structure Engineering in Enhancing Performance of Mosaic Electrocatalysts

Yuting Luo[1], Sum Wai Chiang[2], Lei Tang[1], Zhiyuan Zhang[1], Fengning Yang[1], Qiangmin Yu[1], Baofu Ding[1] & Bilu Liu[1,*]

[1] Shenzhen Geim Graphene Center (SGC), Tsinghua-Berkeley Shenzhen Institute (TBSI) & Tsinghua Shenzhen International Graduate School (TSIGS), Tsinghua University, Shenzhen 518055, P. R. China.

[2] Tsinghua Shenzhen International Graduate School (TSIGS), Tsinghua University, Shenzhen 518055, P. R. China.

Correspondence should be addressed to B.L. (bilu.liu@sz.tsinghua.edu.cn)

**ABSTRACT.**

Understanding the mechanism and developing strategies toward efficient electrocatalysis at gas-liquid-solid interfaces are important yet challenging. In the past decades, researchers have devoted many efforts to improving catalyst activity by modulating electronic properties of catalysts in terms of chemical components and physical features. Here we develop a mosaic catalyst strategy to improve activity of electrocatalysts by engineering their spatial structures. Taking Pt catalyst as an example, the mosaic Pt leads to high catalytic performance, showing a specific activity 11 times higher than uniform Pt films for hydrogen evolution reaction (HER), as well as higher current densities than commercial Pt/C and uniform Pt films. Such a strategy is found to be general to other catalysts (*e.g.*, two-dimensional PtS) and other reactions (*e.g.*, oxygen evolution reaction). The improved catalytic performance of the mosaic catalysts is attributed to enhanced mass transferability and local electric field, both are determined by the occupation ratio of catalysts. Our work shines new light on manipulating electrocatalysis from the perspective of the spatial structure of catalyst, which would guide the design of efficient catalysts for heterogeneous reactions.

## 1. Introduction

Heterogeneous catalytic reactions play central roles in production of chemicals and in energy conversion systems.[1] Much attention has been paid to improving the efficiency of heterogeneous catalysts under electrochemical environments for producing gas fuels, such as hydrogen ($H_2$), oxygen ($O_2$), methane ($CH_4$), and carbon monoxide (CO).[2, 3] For example, the electrochemical hydrogen evolution is considered attractive due to the cleanness and high energy density of hydrogen. Note that these heterogeneous catalytic reactions are usually slow in kinetics, making development of high-efficiency catalysts the key in these fields.[4-6] In recent years, the community has witnessed many successes in developing efficient electrocatalysts, including Pt and other noble metals, metal oxides, chalcogenides, carbides, phosphides, among many others, for the gas-involved electrochemical reactions.[7-14] These works have mostly focused on engineering chemical compositions (*e.g.*, doping, alloying, and phase changing engineering) and/or physical features (*e.g.*, size, shape, and strain engineering) of the catalysts, by which bonding strength between reaction intermediates and catalysts could be modulated so as to improve the catalytic performance.[15-21] For example, Lukowski *et al.* reported that molybdenum disulfide shows a decreased overpotential for hydrogen evolution reaction (HER) when changed from semiconducting 2H phase to metallic 1T phase.[22]. In another work, the electronic structures of atomically dispersed Ru catalysts are precisely modulated by coating metal shells, which exert a proper compressive strain on Ru and resulting in a low overpotential of 220 mV

@ 10 mA $cm^{-2}$ for oxygen evolution reaction (OER).[23] These ways are interesting, while modulating the electronic structures of catalysts via chemical compositions and physical features at the atomic level is challenging. It is therefore interesting to explore a simple strategy to engineer the performance of electrocatalyst. In general, electrochemical process is affected by factors including the chemistry, microscopic structure, geometry of catalysts, *etc.*, which thus can be used to improve catalytic performance.[5, 24-29] This fact gives new opportunities in modulating the performance of electrocatalysts besides chemistry.

Here, instead of focusing on the chemical component and nanostructure of the catalysts, we design a type of mosaic catalyst and demonstrate that the spatial structure of the catalyst is another degree of freedom to modulate performance of electrocatalysts. Taking Pt as an example, we show that mosaic Pt (M-Pt) exhibits a 11-times-higher specific activity for HER than the uniform Pt films, and higher current densities than commercial Pt/C and uniform Pt films at the same overpotentials. Such a mosaic catalyst design strategy has further been extended to two-dimensional (2D) platinum disulphide (PtS) and Ru catalysts for HER and OER, respectively, showing good universality. Experimental and simulation studies show that the fast mass transfer and enhanced local electric field on these mosaic catalysts endow their improved activity. These results indicate that engineering spatial structure of catalysts is an effective way for improving the activity of catalysts.

## 2. Results and Discussion

### Modulating spatial structure of mosaic electrocatalyst for gas-involved reactions

The basic principle of modulating the electrocatalysis by the spatial structures of catalysts is illustrated in Figure 1, where two common factors, mass transfer and the electric field distribution, are considered.

First, the adhesion of gas product/reactant bubbles limits the mass transfer rate to catalysts (Fig. 1a).[20, 30-32] Catalytic sites may be hindered by these bubbles and become inactive, decreasing the efficiency and performance of catalysts if they are not removed efficiently. Second, the strength of local electric field on catalyst would affect the electron energy for catalytic reactions.[28, 33] The electrons taking part in reactions are not only concentrated in the catalyst sites but also in the inactive sites,[34] which would decrease the local operating current density and utilization efficiency of catalyst (Fig. 1a).

By modulating the spatial structure of mosaic catalysts whose catalytic sites and inactive sites are loaded on different regions of matrix, their mass transfer and the electric field distribution can be redesigned (Fig. 1b). The removal efficiency of gas bubbles on the catalyst is mainly controlled by their departure radii ($r_H$) which determines the excess surface energy ($\Delta G^*$) of merged bubbles for detaching the catalyst surface,[35] and is given by,

$$\Delta G^*(r_H) = \sum G_c - \sum G_c' - \sum E_{vis} \qquad (1)$$

where $\sum G_c$, $\sum G_c'$, $\sum E_{vis}$ are the total surface free energies of bubbles before and after coalescence, and total viscous dissipation energy for each bubble, respectively. A high $\Delta G^*$ would be obtained when the $r_H$ of bubbles is small and equal to each other [36]. To this end, the geometrical parameters of the catalytic regions (where bubbles nucleate) and the inactive regions are well redesigned (Fig. 1b), so as to remove bubbles more efficiently and frequently expose the catalytic sites. The strength of local electric field on catalyst could also be redistributed by modulation the spatial structure of catalysts. Electrons with high energy are accumulated to the catalyst sites rather than the inactive sites, which would improve the catalyst use efficiency (Fig. 1b). These effects result in an increase of the thermodynamic driving force for electrochemical reactions and thus an increase of activity on suitably designed catalysts. To identify these mosaic catalysts, the occupation ratio of catalysts ($\Theta_c$) is proposed and is defined by $\Theta_c = A_{catalyst} / A_{electrode}$, where $A_{catalyst}$ and $A_{electrode}$ are the surface areas of catalyst and electrode, respectively. The $\Theta_c$ is equal to 1 for a flat and uniform catalyst (Figure 1a), while it is in the range of 0 to 1 for a mosaic catalyst (Figure 1b).

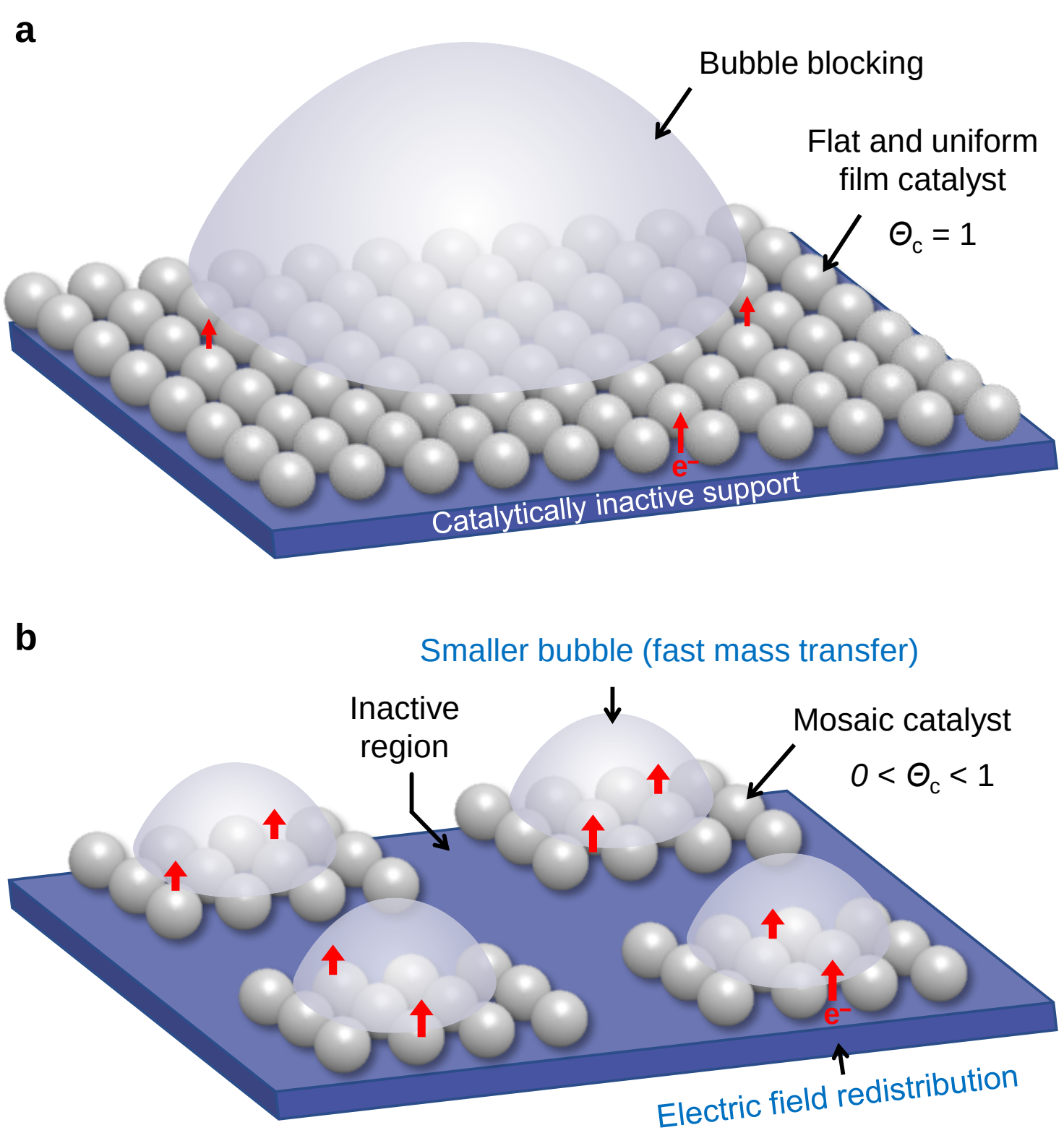

**Figure 1.** Engineering spatial structures of electrocatalysts. (a, b) Schematic showing the adhesion of gas bubbles and the electric field distribution on a flat and uniform film electrode (a) and on a mosaic catalyst (b).

**HER performance of Pt electrocatalysts with spatially structure manipulation**

The advantage of the catalyst with a well-designed spatial structure is tested by Pt catalysts for the HER. Two types of Pt catalysts are fabricated on glassy carbon electrodes, *i.e.*, a uniform and flat Pt film catalyst (U-Pt, Fig. 2a) and a mosaic Pt catalyst with a designed spatial structure (M-Pt, Fig. 2b). The M-Pt consists of square catalyst regions with edge lengths ($L$) of 125 μm and distances between adjacent regions ($D$) of 250 μm. All Pt catalysts are prepared by the same sputtering deposition method (see 'Methods' for details). Atomic force microscopy (AFM) measurements show that the root-mean-square roughness (RMS) of the Pt regions on M-Pt is 1.228 nm, similar to U-Pt (1.216 nm, Fig. S1). The thickness and morphology of M-Pt and U-Pt are kept the same, so that spatial structure is the only difference between the two samples. A Pt foil and a commercial Pt/C film catalyst are also used besides U-Pt as references.

We test the HER performance of these catalysts in a 0.5 M $H_2SO_4$ solution. Typical linear sweep voltammetry (LSV) curves reveal that M-Pt has a high catalytic performance for HER. Specifically, their intrinsic activities are first compared where current densities are determined by geometrical surface areas of Pt ($j_{geo}$).[33, 37] All the samples show a Tafel slope close to 30 mV $dec^{-1}$, indicating that they follow the same rate-determining recombination step (Fig. S2). Interestingly, M-Pt has a smaller overpotential ($\eta$) than the others, reduced by 21 mV (47%) in overpotential than U-Pt at a current density of 10 mA $cm^{-2}$ (Fig. 2c). These results show that M-Pt has a higher catalytic activity than U-Pt though they have identical chemical compositions. At large overpotentials, M-Pt shows a much higher $j_{geo}$ than other samples, with a $j_{geo}$ of 1000 mA $cm^{-2}$ at ~400 mV, which is about twice of the U-Pt (Fig. 2d). Then, we test the electrochemically active surface areas (ECSAs) of these Pt samples using their measured charge of hydrogen desorption peaks after double layer correction. The specific activities ($j_{spe}$) of the samples that are determined by respective ECSAs are compared (Fig. 2e). The results show that the $j_{spe}$ of M-Pt is 16.7 mA $cm^{-2}$ at $\eta$ = 50 mV, about one order of magnitude higher than U-Pt and compare favorably to the Pt foils and commercial 20 wt% Pt/C catalysts (Fig. 2e and Table S2).

We further analyze the catalytic performance of the Pt catalysts at different current densities. The value of $\Delta\eta/\Delta\log|j|$ ($R_{\eta/j}$, defined as overpotential $\eta$ divided by current density $j$), a recently proposed indicator to evaluate the performance of catalysts at different current densities,[20] is used to evaluate how much overpotential is needed as the current density increases (Fig. 2f). The $R_{\eta/j}$ of M-Pt remains small (414 mV $dec^{-1}$) but becomes large for U-Pt and Pt foil (1015 mV $dec^{-1}$) as the current density increases to 1000 mA $cm^{-2}$, indicating that M-Pt maintains its performance at high current densities. We also compare the mass transfer on U-Pt and M-Pt at different current densities (Fig. 2g). Although the densities of $H_2$ bubbles on both catalysts increase with the current density, M-Pt shows two orders of magnitude greater bubble density than that of U-Pt. At a small current density of 10 mA $cm^{-2}$, the majority of $H_2$ bubbles on both samples have diameters smaller than 250 μm, which is close to $D$ (distance between adjacent Pt regions in M-Pt catalyst). As current density increases, more active sites are blocked by the grown bubbles on U-Pt, while the sizes of bubbles on M-Pt remain less than $D$ due to high bubble removal efficiency (Fig. 2h). *In situ* optical microscopy (OM) observations show that bubbles on M-Pt remove at a high rate in a 'bubble relay' mode, whereas bubbles on U-Pt are randomly generated and removed in a slow rate (Movie S1). According to the formula (1), the higher density and the smaller size of hydrogen bubbles (*i.e.*, smaller $r_H$) on M-Pt than on U-Pt promote the mass transfer efficiency of $H_2$ on M-Pt. Overall, these results show that M-Pt has an improved HER performance.

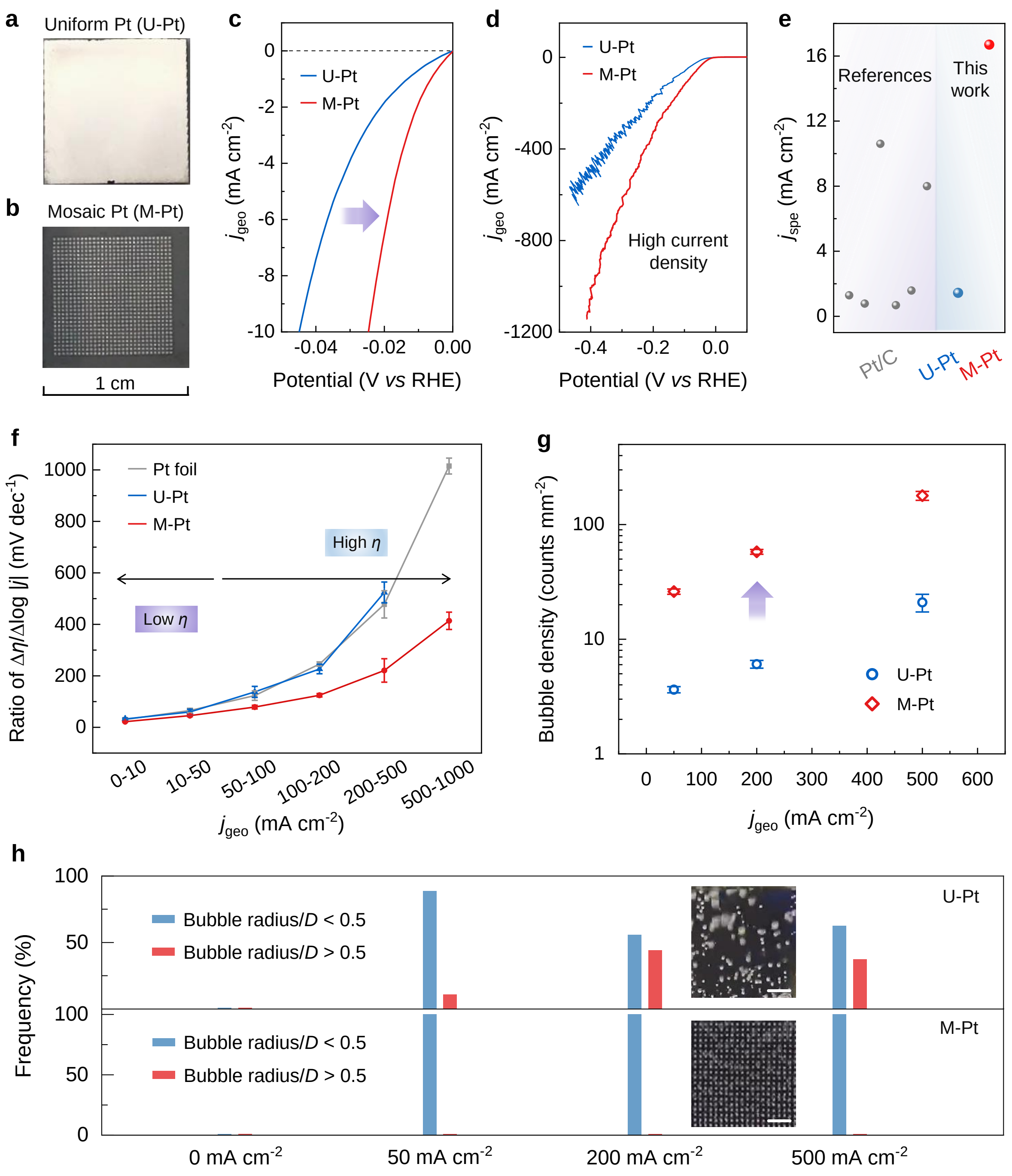

**Figure 2.** HER performance of different Pt catalysts. (a,b) Photos showing (a) an uniform and flat film Pt catalyst (U-Pt) and (b) a mosaic Pt catalyst with its spatial structure well-designed (M-Pt) on glassy carbon electrodes. (c) Linear sweep voltammetry (LSV) curves of two types of Pt catalysts, showing their geometric current densities ($j_{geo}$) and revealing a better intrinsic activity of M-Pt than the U-Pt. (d) LSV curves and $j_{geo}$ of Pt catalysts at large overpotentials. (e) Specific activities ($j_{spe}$) of several Pt

samples at overpotential of 50 mV. (f) Values of $\Delta\eta/\Delta\log|j|$ ($R_{\eta/j}$) for Pt catalysts in different current density ranges. All points are measured three times, and error bars correspond to the standard deviations. Densities (g) and diameters (h) of $H_2$ bubbles on U-Pt and M-Pt at different current densities. Here *D* is the distance between adjacent Pt regions in M-Pt. All points are tested three times, and error bars correspond to standard deviations.

**Universality of catalyst design to improving catalytic performance**

As a demonstration of the universality of such a catalyst design strategy, we extend it to other catalysts and gas-involved reactions. The first example is PtS for the HER. The PtS is grown by chemical vapor deposition (CVD) into either uniform film catalyst (U-PtS) or mosaic catalyst with the redesigned triangle shaped patterns (Mt-PtS, Figs. 3a and S3, see 'Method' for details). We find that the Mt-PtS shows a much better HER performance than U-PtS. For example, Mt-PtS shows a $j_{geo}$ = 3329 mA $cm^{-2}$ at $\eta$ = 650 mV (Fig. 3b), while U-PtS only delivers 305 mA $cm^{-2}$.

The second example is Ru catalyst for the OER. Three Ru catalysts are prepared, including two mosaic catalysts with either triangle or square patterns (a M-Ru and a Mt-Ru), and one uniform and flat film catalyst (U-Ru, Fig. 3c and 3d, see 'Methods' for details). The polarization curves show that the M-Ru and Mt-Ru have better OER performance than U-Ru (Fig. 3e). For example, the Mt-Ru shows a $j_{geo}$ = 183 mA $cm^{-2}$ at $\eta$ = 320 mV, which is three times higher than that of U-Ru at the same overpotential. These two examples show that the catalyst design strategy is general to other materials (PtS) and other reactions (OER), besides Pt for HER.

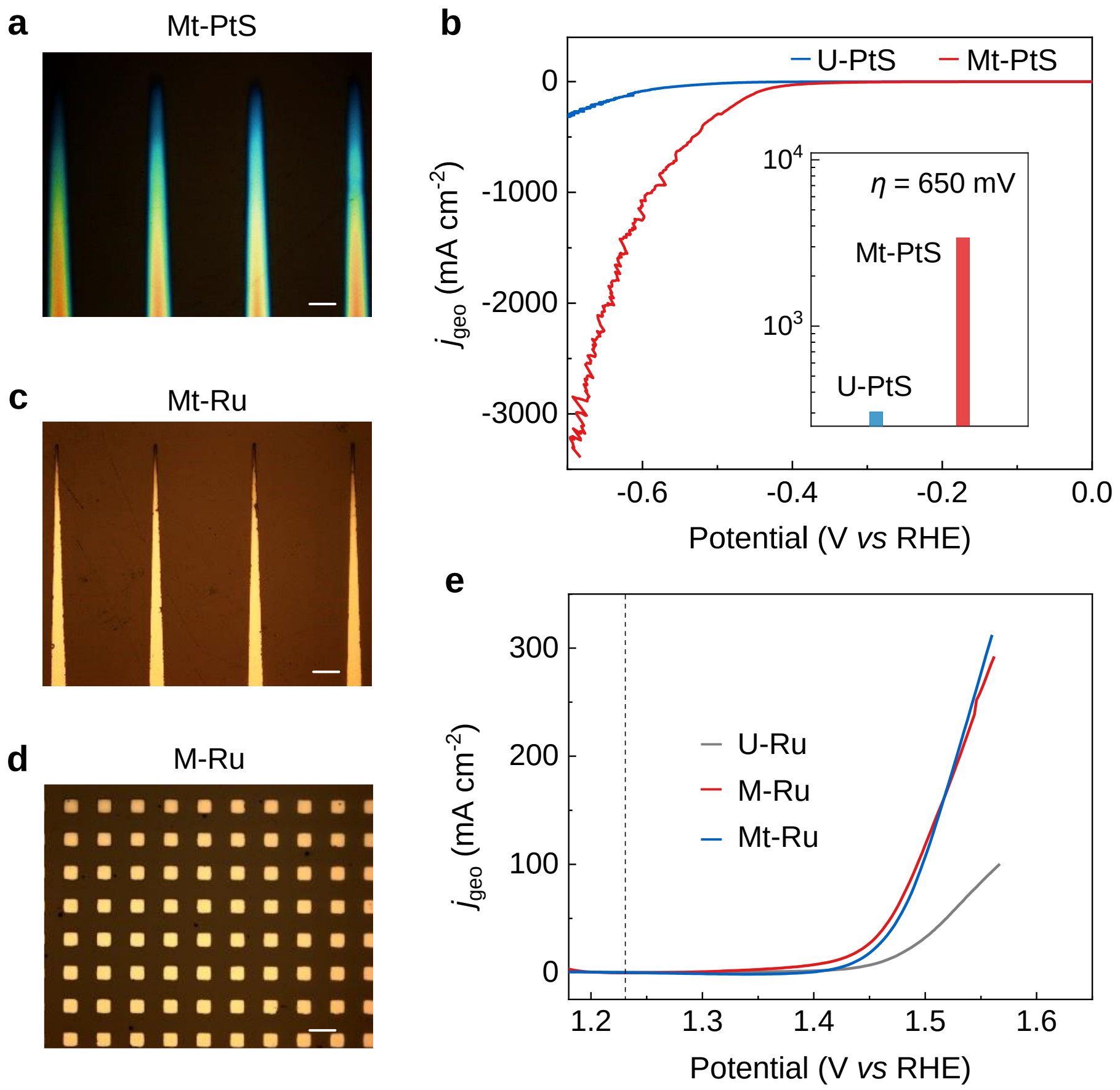

**Figure 3.** Universality of the catalyst design strategy for different catalysts and reactions. (a) An optical microscopy image of Mt-PtS catalyst. (b) LSV curves of U-PtS and Mt-PtS for HER in 0.5 M $H_2SO_4$. (c, d) Optical microscopy images of (c) Mt-Ru and (d) M-Ru catalysts. (e) LSV curves of U-Ru, M-Ru, and Mt-Ru catalysts for oxygen evolution reaction in 0.5 M $H_2SO_4$. All scale bars are 200 µm.

**Tuning catalytic performance of catalysts by their engineering spatial structures**

We then show how the geometrical features of catalyst affect their performance. Six mosaic Pt catalysts (namely M1-Pt, M2-Pt, M3-Pt, M4-Pt, M5-Pt, and M6-Pt) with sequentially decreased $\Theta_c$ are fabricated (Fig. 4a and Table S1) for HER tests. From their LSV curves, all the M-Pt catalysts show better HER performance than Pt foil and the U-Pt. Among them, M5-Pt shows the best performance with $\eta$ = 13 mV at $j_{geo}$ = 10 mA cm$^{-2}$ and $\eta$ = 98 mV at $j_{geo}$ = 1000 mA cm$^{-2}$ (Fig. 4b). In contrast, M1-Pt, M6-Pt, and Pt foil exhibit large overpotentials of 25 mV, 18 mV, and 46 mV at $j_{geo}$ = 10 mA cm$^{-2}$ as well as 396 mV, 175 mV, and 671 mV at $j_{geo}$ = 1000 mA cm$^{-2}$, respectively (Figs. S2 and S4). We further analyze the $R_{\eta/j}$ ratios of these Pt catalysts (Fig. 4c). With increasing current density, M1-Pt and M6-Pt have large $R_{\eta/j}$ ratios of more than 200 mV dec$^{-1}$ while M5-Pt has a small $R_{\eta/j}$ ratio of less than 100 mV dec$^{-1}$. All these results indicate that engineering the spatial structures of catalysts is an easy and effective way to improve heterogeneous catalytic performance.

The parameter, $\Theta_c$, is used to describe the spatial structure of the catalysts. By summarizing the overpotentials at both low (10 mA cm$^{-2}$, *i.e.*, $\eta_{10}$) and high $j_{geo}$ (1000 mA cm$^{-2}$, *i.e.*, $\eta_{1000}$) of samples and their different $\Theta_c$ values, we study how factors influence their catalytic performance, as shown in Figs. 4d-e and S5. The results indicate that both $\eta_{10}$ and $\eta_{1000}$ show the correlations with $\Theta_c$ values (Fig. 4d-e). At smaller $\Theta_c$ values, the $\eta_{10}$ reduces that means a better catalytic performance. Moreover, the correlation between overpotential and $\Theta_c$ value is not a simple one. In the case of 1000 mA cm$^{-2}$, the approximately linear correlation between overpotential and $\Theta_c$ value is not as clear as low current density (Fig. 4e). Such a deviation from the correlation can be attributed to great contribution of mass transfer at high current densities. Relationships between other structural parameters of Pt catalysts and their catalytic performance are also studied. The results show no obvious correlations between catalytic activity and wettability of electrodes (Fig. S6) as well as the ratio of catalyst perimeter to area, meaning the edges of Pt regions contribute negligibly to the HER activity of M-Pt (Fig. S7). The above results indicate that a properly low $\Theta_c$ increases the efficiency of each catalyst region and improve the performance of electrocatalysts.

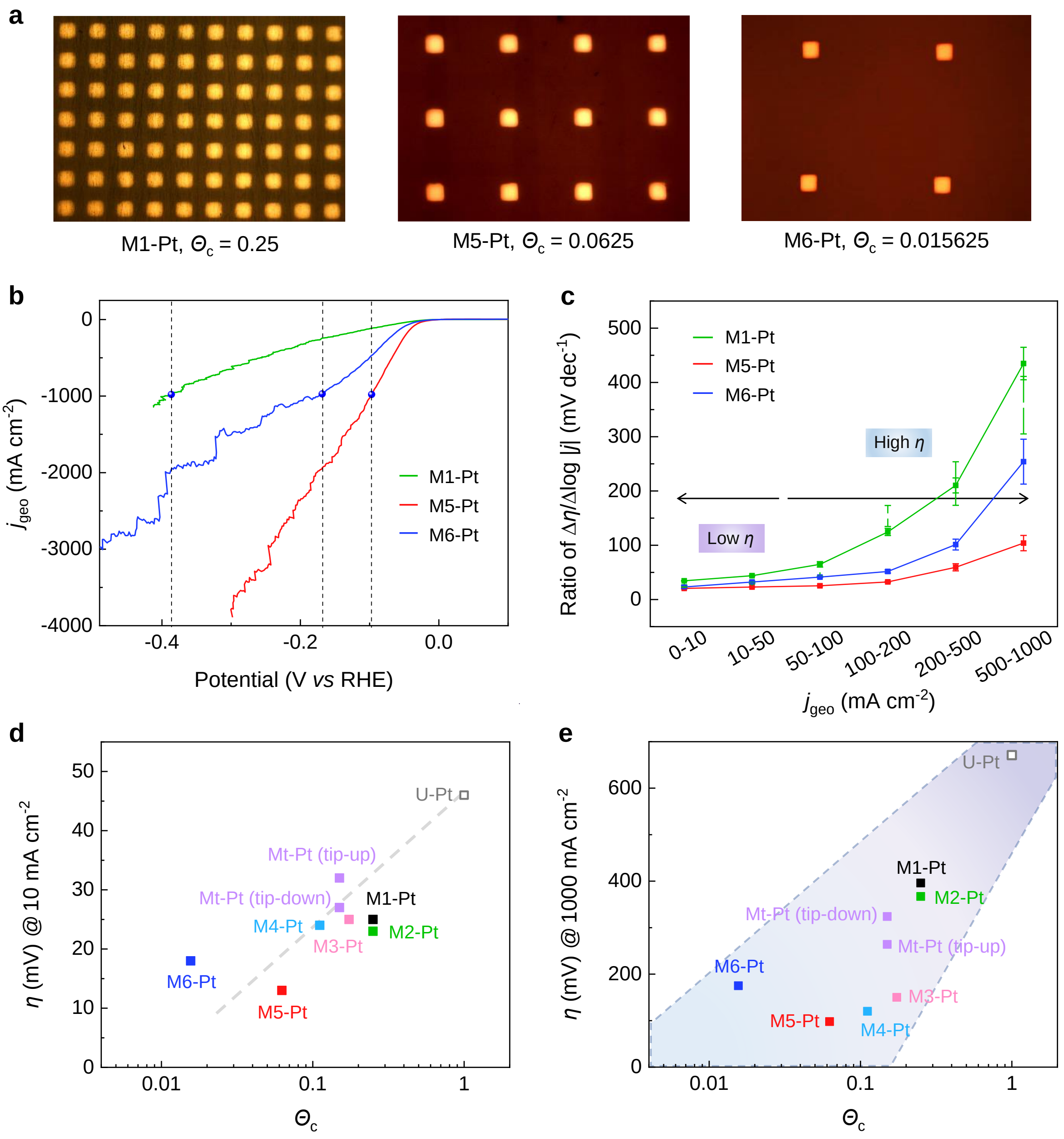

**Figure 4.** Tuning HER performance of mosaic Pt catalysts by their engineering spatial structures. (a) Optical microscopy images showing three Pt catalysts with different $\Theta_c$, marked as M1-Pt ($\Theta_c = 0.25$), M5-Pt ($\Theta_c$ = 0.0625), and M6-Pt ($\Theta_c$ = 0.015625). (b) LSV curves of the three representative Pt catalysts. (c) $R_{\eta/j}$ for the three Pt catalysts in different current density ranges. All points were tested three times, and the error bars correspond to the standard deviations. (d, e) Summaries of (d) the $\eta_{10}$ (e) the $\eta_{1000}$ values of Pt catalysts with different $\Theta_c$ values, where the $\eta_{10}$ and $\eta_{1000}$ are defined as overpotentials at $j_{geo}$ = 10 mA cm$^{-2}$ and $j_{geo}$ = 1000 mA cm$^{-2}$, respectively).

**Effect of size of $H_2$ bubbles on mass transfer and HER performance**

Having verified the effectiveness of modulating spatial structure of catalysts to improve their activity, we now explore the contribution of bubble size to mass transfer and catalytic performance. Two opposite placed orientations of triangle shape Pt catalysts (Mt-Pt) are used with the tips of triangle regions up or down referring to the direction of gas flow, *i.e.*, tip-up and tip-down (Fig. 5a and Fig. S8). The two samples show the same chemical composition and morphology, making the orientation is the only different parameter. As a result, the two samples show the opposite directions of the Laplace forces on bubble caused by the asymmetrically triangle shapes of the catalyst regions (Fig. S9),[38] and thus the effect of mass transfer on catalytic performance can be independently studied. As shown in Figure 5a, both samples show the same catalytic performance at current densities < 200 mA $cm^{-2}$, while the tip-up sample shows better performance than the tip-down one at current densities > 200 mA $cm^{-2}$ (Movie S2). As current densities get higher, their HER difference becomes more significant. For example, the overpotential at $j_{geo}$ = 1000 mA $cm^{-2}$ decrease by ~27% for tip-up Mt-Pt compared to tip-down Mt-Pt. We also verify the role of mass transfer on high-current-density HER by intentionally blocking the removal of $H_2$ bubbles with a piece of aerophilic polytetrafluoroethylene on M-Pt catalyst (Fig. S10). These results show that the mass transfer rate plays a key role in determining the catalytic performance of catalysts are high current densities.

The square of departure radii of bubbles, $r_{\mathrm{H}}^2$, is related to blocked areas of the catalysts following the Cassie-Baxter mode (see 'Methods' for details), which is used to evaluate the critical radius of $H_2$ bubbles that influenced mass transfer ability (Figs. 5b-c and S8, Movie S2). The radii of bubbles are obtained by analyzing tens of $H_2$ bubbles (Fig. S11). The results show that the $r_{\mathrm{H}}^2$ of $H_2$ bubbles on tip-down Mt-Pt is larger than that on the tip-up one (Fig. 5d). As shown in Fig. 5a, the current density at critical point is about $j_{geo}$ = 200 mA cm$^{-2}$, referring to a critical $r_H^2$ of ~$10^4$ μm$^2$ and an average bubble radius of 100 μm. Oversized $H_2$ bubbles will impede mass transfer on the catalyst and increase the mass-transfer overpotential of the whole reaction system, resulting in low catalytic performance.[33] Taking together, these results show the importance of mass transfer on catalytic performance, especially at high current densities.

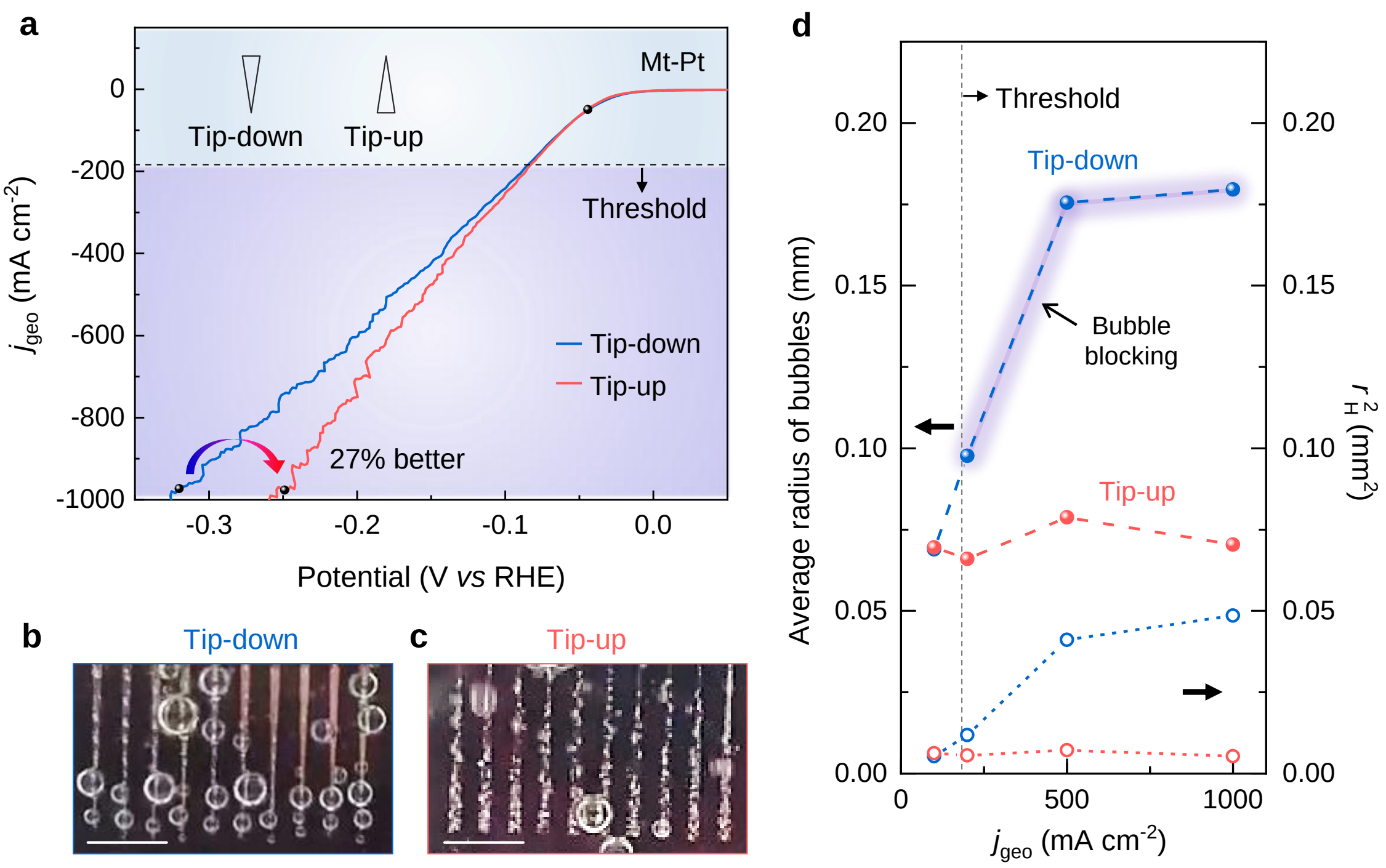

**Figure 5.** Effect of bubble size on mass transfer and HER performance. (a) LSV curves of Mt-Pt with tip-up (red) and tip-down (blue) orientations. (b, c) Photos showing different transfer ability of $H_2$ bubbles on (b) tip-down and (c) tip-up samples. By changing catalyst orientations, their performance changes due to different mass transfer abilities. Scale bars in (b) and (c) are 2 mm. (d) Relationships between average radius (left Y-axis) and $r_H^2$ (right Y-axis) of $H_2$ bubbles to HER performance, indicating a critical bubble radius ~100 μm, over which $H_2$ bubbles impede the mass transfer process in Mt-Pt samples.

## Electric field redistribution

Besides mass transfer, we study the electric field redistributions of different M-Pt electrocatalysts by the finite element analysis. As a reference, U-Pt is also studied. First, local electric field distribution is explored. The results show that electric field is uniformly distributed on the U-Pt with a strength of $E_0$ (Fig. 6a). As for M-Pt, the Pt regions separated by catalytically inactive carbon can act as ‘point sinks’ of the electrical potential, greatly enhancing the strength of electric field closed to these Pt regions (Fig. 6b). As a result, the electrical energy transport is congesting into the electrode through these small Pt regions and causes the electric field strengthening localized at their vicinity. This enhancement would result in reaction improvement on M-Pt by providing more electrical energy to the reactions. Moreover, the strength of the electric field is sensitive to the coverage and exposure of M-Pt (Fig. 6c). For the three-phase interface involved reactions such as HER, we also study the effect of gas bubbles on the distribution of electric field and the results indicate that the strength of local electric field can be further enhanced by the existence of gas bubbles (Fig. 6d and Fig. S12). Note that the effect of mass transfer is not considered in the simulation as the dynamic process is usually complicated. In addition to the local electric field distribution, we have also considered the maximum achieved electric field strength, $E_{max}$, near the Pt regions in different M-Pt and the U-Pt. A ratio of $E_{max}/E_0$ is used as an indicator to describe how strong the electrical field is in each catalyst. We plot the values of $E_{max}/E_0$ of different M-Pt with their corresponding $\Theta_c$ (*i.e.*, $L^2/D^2$) for the square shape catalysts (see ‘Method’ for details). The results show that the magnitude of $E_{max}/E_0$ of M-Pt can be as high as a value of tens level (Fig. 6e). In theory, the increase of $E_{max}$ will improve local reaction significantly.[28, 33] Therefore, our simulation results show that a smaller $\Theta_c$ will lead to a higher $E_{max}$.

Furthermore, the average value of electric field strength over the electrode area (including areas of Pt regions and areas of inactive regions, $E_{ave}$) are analyzed to study geometric activities of catalysts based on the electrode areas. We find that $E_{ave}/E_0$ reaches the peak when the $\Theta_c$ is in range from 0.1 to 0.3 (Fig. 6e). The enhancement of the reactions therefore should not be linear and M-Pt with a proper occupation of catalyst may have a high electrode activity for reactions, giving opportunities to future engineering of the spatial structures of catalysts. Note that the optimized ranges for obtaining the best electrode activity will change by the geometrical parameters and microscopic morphologies of each catalyst regions. Here, the electrode current density determined by the project area of electrode ($j_{electrode}$) is used for comparison as it is an index for practical use. Guided by this feature, we find optimized M-Pt samples (with $\Theta_c$ of 0.1736 or 0.1111) show overpotentials much smaller than those of U-Pt and commercial Pt/C film catalyst at $j_{electrode}$ =100 mA $cm^{-2}$ (Figs. 6f and S13). This design strategy is particularly meaningful for expensive catalyst. Together, our simulation results show that the improved catalytic activity of M-Pt is ascribed to the strengthening of local electric field.

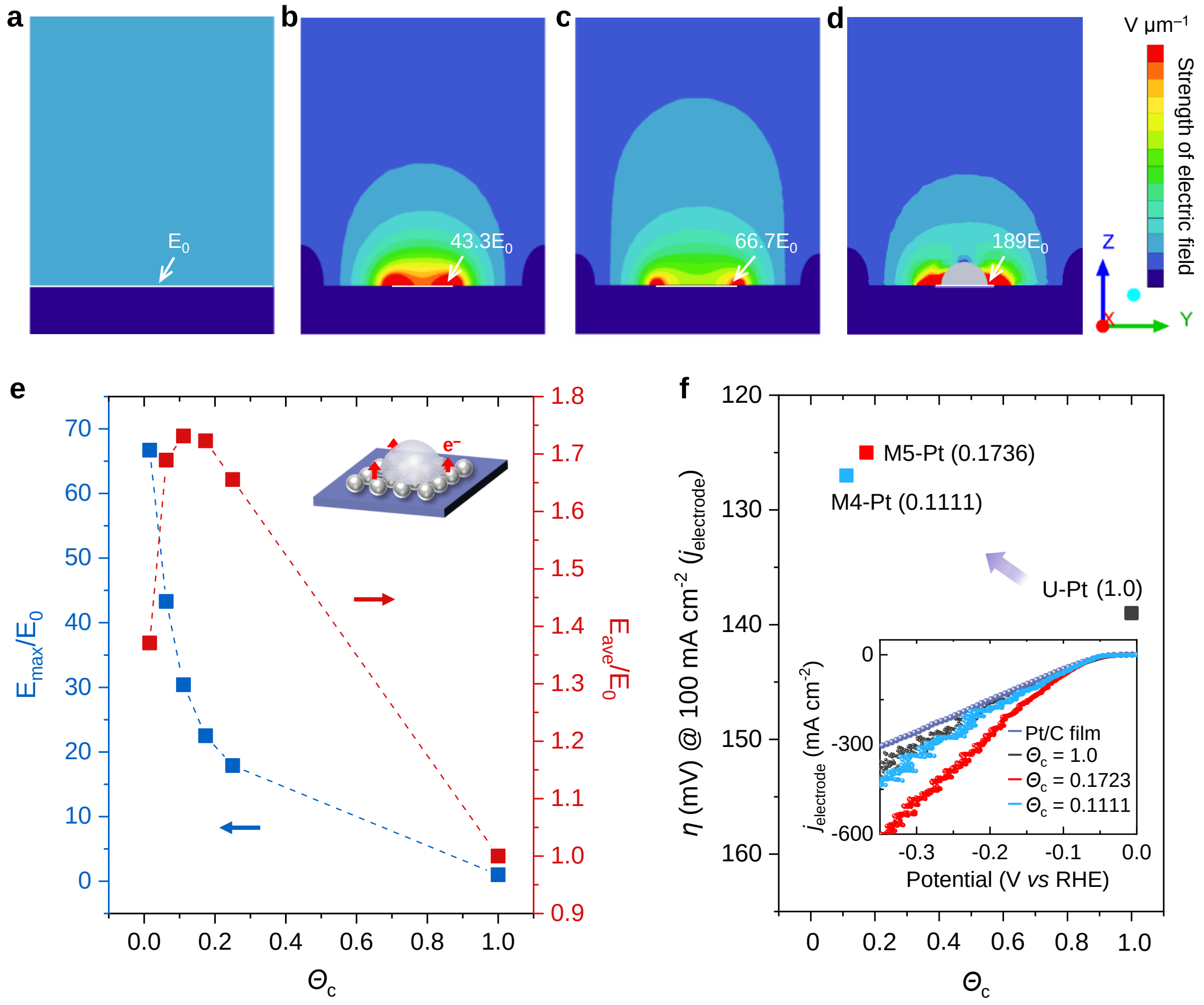

**Figure 6.** Theoretical analysis. (a-d) Local electric field distribution in electrolyte in different catalysts, including (a) U-Pt, (b) M-Pt with a $\Theta_c = 0.0625$, (c) M-Pt with a $\Theta_c = 0.1111$, and (d) a hemispheric gray bubble with a diameter of 100 μm sit at the center of M-Pt with a $\Theta_c = 0.0625$. The strength of the electric field on U-Pt (denoted as light blue color) is $E_0$ and set as the reference. The Pt regions are shown by white lines. (e) The correlations between $\Theta_c$ of the M-Pt and their $E_{max}/E_0$ and $E_{ave}/E_0$, where $E_{max}$ and $E_{ave}$ are the maximum electric field and average electric field based on electrode surface areas. (f) M-Pt samples show overpotentials comparable to those of the U-Pt and commercial 20 wt% Pt/C film catalyst at electrode current density ($j_{electrode}$) of 100 mA $cm^{-2}$. The inset shows their polarization curves.

## 3. Conclusions

We have demonstrated a mosaic catalyst for efficient electrocatalysis, based on a new degree of freedom of catalyst design, *i.e.*, their spatial structures.. Two key considerations in such a catalyst design are the local electric field strength and mass transfer ability, both been affected by the spatial structures of the catalyst. Besides the HER and OER reported here, this catalyst design may have universal relevance to various catalytic reactions involving gas-involved heterogeneous reactions, including carbon dioxide electrochemical reduction, nitrogen electrochemical reduction, and oxygen reduction reaction, because mass transfer rate and electric field redistribution are common features in these processes.

**Supporting Information**

Supporting Information is available from the Wiley Online Library or from the author.

**Acknowledgements**

We thank Mr. Rongjie Zhang for assistance in e-beam metal deposition. We acknowledge support from the NSFC (Nos. 51722206 and 51991343), the Youth 1000-Talent Program of China, Guangdong Innovative and Entrepreneurial Research Team Program (No. 2017ZT07C341), the Bureau of Industry and Information Technology of Shenzhen for the "2017 Graphene Manufacturing Innovation Center Project" (No. 201901171523), and the Shenzhen Basic Research Project (JCYJ20190809180605522).

**Conflict of Interest**

The authors declare no conflict of interest.

**Keywords**

Electrocatalysis, gas-liquid-solid interface, catalysts, hydrogen evolution, catalytic activity.